# DinoMUSE: A DINOv2-guided Framework for Unsupervised Lateral Motion Estimation in Ultrasound Elastography

Xiyao Meng, Xingyue Wei, Baihua Luo, Bo Peng, Jianwen Luo, *Senior Member, IEEE*

***Abstract*****—Quasi-static ultrasound elastography (USE) is a non-invasive imaging method for assessing soft tissue mechanical properties. Limited by low resolution, low sampling frequency and lack of phase information in the lateral direction, accurate lateral motion estimation remains a critical bottleneck in USE. Existing methods mainly rely on local image matching and physical constraints (such as tissue incompressibility assumption and effective Poisson's ratio), which lack sufficient contextual information. We present DinoMUSE, a DINO-guided mixture-of-experts (MoE) framework for unsupervised lateral motion estimation in USE. DinoMUSE incorporates high-level visual representations learned by DINOv2 into the motion estimation network to provide global contextual guidance. To facilitate effective interaction between DINOv2 features and local matching features, an MoE-based feature-fusion module adaptively transforms and integrates the two representations. Evaluations against NCC, GLUE, RAFT, and sPICTURE on simulation and *in-vivo* breast datasets show that DinoMUSE substantially improves lateral estimation accuracy while maintaining competitive axial performance. Specifically, DinoMUSE reduces the mean square error (MSE) of lateral displacement and strain estimation by 60% and 65%, respectively, compared with a SOTA algorithm on the simulation dataset. For *in-vivo* breast data, DinoMUSE achieves axial strain SNR and CNR of 5.70 and 4.82, and lateral strain SNR and CNR of 4.61 and 2.23, compared to 4.11 and 3.32, and 4.57 and 2.13 for the best competing method, alongside improved lesion visualization. These findings indicate the potential of leveraging transferable visual representations and adaptive feature fusion to improve lateral motion estimation in USE. The code is available at xxxxxxxx.**

This work was supported in part by the National Natural Science Foundation of China (NSFC) under Grants 62531004, 62027901, and 82372561, the Beijing Natural Science Foundation under Grants 7252066 and L2611016, and the Capital's Funds for Health Improvement and Research under Grant 2026-2-2025. (Xiyao Meng and Xingyue Wei contribute equally to this work.) (Corresponding author: Jianwen Luo.)

Xiyao Meng, Xingyue Wei, Baihua Luo, and Jianwen Luo are with the School of Biomedical Engineering, Tsinghua Medicine, Tsinghua University, Beijing 100084, China (e-mail: luo_jianwen@tsinghua.edu.cn).

Bo Peng is with the School of Computer Science, Southwest Petroleum University, Chengdu, Sichuan 610000, China.

## I. Introduction

Ultrasound elastography (USE) is a non-invasive imaging technique for assessing soft-tissue stiffness. In quasi-static USE, tissue motion (i.e., displacement) induced by external compression or physiological pulsation is tracked across consecutive ultrasound radio-frequency (RF) frames. By estimating axial and lateral displacement fields between these frames, strain fields can be derived to reveal mechanical differences among various tissues [1]. Unlike conventional B-mode ultrasound imaging that only reflects anatomical information, USE provides stiffness-related insights. It has been widely applied in many clinical scenarios, such as benign-malignancy screening for breast [2], thyroid nodules [3], and so on.

Accurate estimation of lateral displacement and strain provides essential complementary information to axial-only measurements for reconstructing full strain-tensor fields, calculating Poisson's ratio, and characterizing heterogeneous pathological tissues [4-7]. Despite high-quality axial (i.e., along the ultrasound insonification) motion estimation, lateral (i.e., in plane and perpendicular to the ultrasound insonification) motion estimation remains a bottleneck [8], due to the intrinsic properties of ultrasound imaging including lower sampling frequency (i.e., lower line density) [9], low spatial resolution [10], and lack of RF phase information (i.e., oscillations in the point spread function) in the lateral direction [11]. Multiple beamforming approaches have been explored to address these issues, including spatial angular compounding [12, 13], synthetic aperture [10] and transverse-oscillation beamforming [11, 14]. In addition, some studies have adopted the tissue incompressibility assumption [15-17] or other physical losses [5, 6] to infer lateral information from axial measurements. Although these approaches have improved the estimation of lateral motion, their performance remains limited. Moreover, existing methods predominantly rely on local image matching to estimate motion, which cannot provide sufficient contextual information for resolving ambiguous correspondences in challenging regions.

This motivates the incorporation of high-level visual representations that can provide broader contextual cues to complement local matching information. Self-supervised visual foundation models, such as DINOv2 [18], have exhibited strong transferable representation capabilities across diverse medical-imaging tasks [19-21]. Their ability to capture global semantic and contextual information [18, 20] can provide priors to facilitate lateral motion estimation.

Motivated by the above observations, this paper presents DinoMUSE, a DINOv2-guided mixture-of-experts (MoE) framework for unsupervised lateral motion estimation in USE. Specifically, we embed the rich global representations learned by DINOv2 into the motion estimation network, leveraging their high-level visual context to guide lateral motion estimation. However, these global representations differ fundamentally from the local matching features learned by the motion estimation network, creating a representation gap that can hinder their effective interaction. To bridge this gap, we further develop an MoE-based feature fusion module that adaptively transforms and integrates the two representations for lateral motion estimation.

The main contributions of this paper are as follows:

(1) We propose an unsupervised DinoMUSE framework that incorporates rich global representations from DINOv2 into the motion estimation network, providing high-level visual guidance for lateral motion estimation.

(2) We develop an MoE-based feature fusion module to address the representation gap between DINOv2 global features and the local matching features learned by the motion estimation network.

(3) We conduct extensive evaluations on simulations and *in-vivo* breast datasets, demonstrating consistent improvements in lateral displacement and strain estimation while maintaining competitive axial performance.

## II. Related Work

### A. Motion Estimation Algorithms in Elastography

Motion estimation approaches in USE are broadly divided into traditional and deep-learning-based methods.

Traditional methods include window-based techniques and global regularized optimization. Represented by normalized cross-correlation (NCC) [22, 23], window-based methods track motion via local speckle similarity. They deliver acceptable accuracy with moderate noise robustness [24] yet suffer from performance degradation under large deformation due to speckle decorrelation [25] and involve high computational complexity [22, 26]. A re-correlation post-processing step is commonly adopted for NCC to achieve sub-sample displacement refinement[27, 28]. Global-optimization methods such as GLUE [29], OVERWIND [30], and SOUL [31] solve displacement fields by minimizing intensity and continuity cost functions. They are more robust to speckle decorrelation and out-of-plane motion [32] and yield higher signal-to-noise ratio (SNR) and contrast-to-noise ratio (CNR) [29]. However, uniform global smoothing tends to blur subtle lesion boundaries. Moreover, heavy iterative computation and time-consuming hyper-parameter tuning limit clinical real-time use.

Deep-learning methods fall into supervised and unsupervised categories. Supervised models (e.g., MPWCNet variants [32], GLUE-Net [33]) require finite-element or phantom-simulated ground-truth displacements. Their clinical translation is limited by scarce annotated ultrasound data [34]. Unsupervised motion estimation networks (e.g., LiteFlowNet [35], MPWCNet++ [36], ReUSENet [37] and MF-UCNN [38]) eliminate ground-truth requirements and better fit unlabeled clinical data.

To improve lateral motion estimation, early work derived lateral strain from axial results under the tissue-incompressibility assumption [15-17] to avoid direct lateral estimation. However, this assumption introduces obvious biases for heterogeneous tissues such as tumors, muscles and cartilages [39]. Other representative traditional solutions include Bayesian regularization [40], lateral interpolation [41-43], effective Poisson's ratio (EPR)-driven constraints [44], poroelastography-based PULSE formulation built upon biphasic tissue physics [45], regularization-based full 2-D strain-tensor reconstruction [46], and MixTURE deploying mixed-order second-derivative penalties [7]. Nevertheless, these approaches still rely on manual hyper-parameter tuning and suffer from slow inference speed. Recently, some deep networks also embed incompressibility priors [47]. Besides, ElastNet leverages a dual-network physics-informed neural network (PINN) formulation governed by static elastic equilibrium partial differential equations, which infers latent lateral displacement purely from measured axial displacement inputs [48]. sPICTURE uses EPR constraints to exploit axial-lateral mechanical coupling [5]. RAFT-USENet adopts strain-compatibility losses and decoupled separable GRU modules [6], but relies on simulation-driven supervised pre-training. SMURF adapts RAFT [49] for self-supervised B-mode elastography and achieves competitive lateral estimation without simulation-based ground-truth labels [50].

However, these CNN-driven methods depend heavily on local features and lack global context, causing global deformation drift for complex cases with multiple lesions or large lesions.

### B. DINOv2 in Medical-Image Analysis

Self-supervised learning mitigates label scarcity in medical image analysis [51]. Within this paradigm, DINOv2 leverages self-distillation on massive natural image datasets to learn general-purpose and transferable visual representations without manual annotations [18]. DINOv2 yields versatile visual features for both image-level and pixel-level downstream tasks. In particular, its patch-level token representations enable semantic segmentation, as well as dense and sparse cross-image visual matching [18]. These properties have motivated the adoption of DINOv2 features as visual priors for downstream medical image analysis, including segmentation, classification, and registration [20, 52-55].

For ultrasound motion estimation, DINOv2 exhibits three particular strengths. First, multi-head self-attention enables joint global-local encoding, capturing long-range anatomical dependencies while preserving local speckle textures. Second, unlike UNet-based encoders, frozen DINOv2 generates effective ultrasound representations without fine-tuning on elastography labels, greatly alleviating the scarcity of medical annotations. Third, self-distillation yields deformation-robust patch embeddings. Feature similarity of lesion-related patches remains stable under large deformations [20].

Therefore, this work integrates DINOv2 features into a motion estimation network, leveraging their rich semantic information and ability to establish dense correspondence to improve lateral motion estimation in USE.

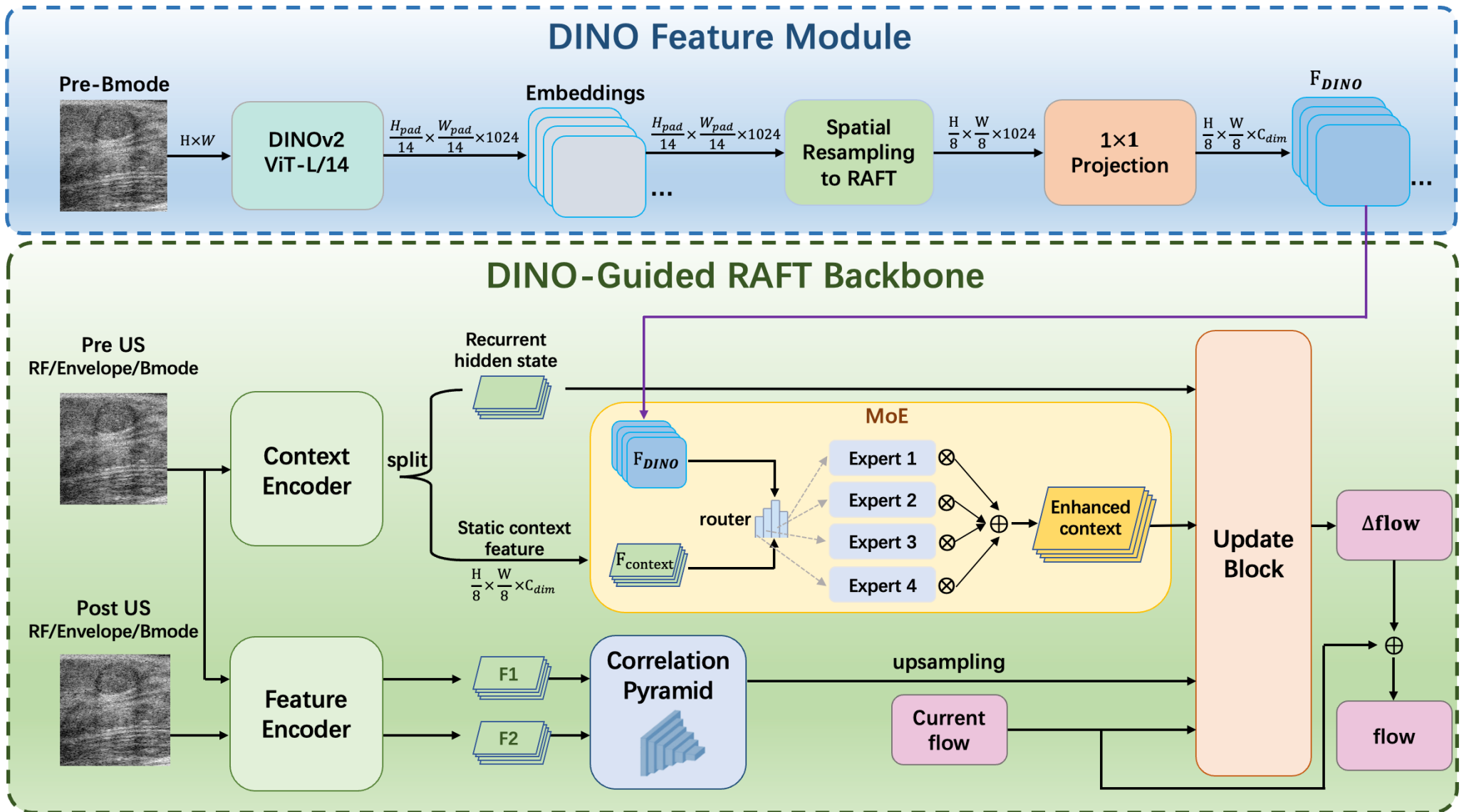

Fig. 1. Pipeline of the proposed DinoMUSE. It extracts global semantic representations for pre-compression B-mode images via frozen DINOv2, and adopts an MoE-based feature fusion module to inject DINOv2 features into the RAFT backbone for unsupervised ultrasound displacement estimation.

### C. Mixture of Experts

MoE enhances model capacity by introducing multiple expert sub-networks, with a gating router dynamically selecting or weighting the experts based on the input [56].

MoE has been widely adopted in medical image analysis to mitigate task conflicts and negative transfer under multi-task and multi-modal scenarios [57-60]. MoE-Morph deploys morphology-aware expert branches to handle diverse anatomical deformations for medical registration [59]. MedSegX proposes an adapter-based contextual MoAE for task-aware parameter decoupling, reducing performance degradation caused by simple multi-source data aggregation in open-world medical segmentation [60].

Motivated by these studies, this paper proposes an MoE-based feature fusion module to adaptively integrate DINOv2 features with local matching features learned by the motion estimation network.

## III. Methods

In this section, we present the technical details of the proposed DinoMUSE. We begin with a task definition of USE in Section III-A. The DINOv2 feature extraction process is described in Section III-B. Section III-C presents the MoE-based feature fusion module. Finally, we describe the loss function in Section III-D.

### A. Task Definition of Ultrasound Elastography

In quasi-static USE, two frames of RF data are acquired before and after tissue compression. The displacement field is then estimated to compute the strain field. We denote the 2-D dense displacement field as $\boldsymbol{u} = [u_a, u_l]^T$, where $u_a$ and $u_l$ denote the axial and lateral displacements, respectively.

Under the small-deformation assumption, the 2-D strain tensor is derived from the gradient of the displacement field. Axial strain $\varepsilon_{aa}$ serves as a well-established diagnostic metric whereas lateral strain $\varepsilon_{ll}$ provides complementary mechanical information about soft tissues. The shear strain $\varepsilon_{al}$ further shows mechanical differences along lesion boundaries [61]:

$$\varepsilon_{aa} = \frac{\partial u_a}{\partial a}, \qquad \varepsilon_{ll} = \frac{\partial u_l}{\partial l}, \qquad \varepsilon_{al} = \frac{1}{2}\left(\frac{\partial u_a}{\partial l} + \frac{\partial u_l}{\partial a}\right) \tag{1}$$

$\varepsilon_{al}$ is composed of two terms: the axial shear strain $\frac{\partial u_a}{\partial l}$ and lateral shear strain $\frac{\partial u_l}{\partial a}$[7].

### B. DINOv2 Feature-Extraction and Alignment Module

Taking RAFT as the motion estimation backbone, we introduce the frozen DINOv2 foundation model to extract global semantic features from B-mode images and inject these features into the context-encoding branch of RAFT to improve lateral motion estimation.

The extraction pipeline for DINOv2 features is shown in Fig. 1. Zero-padding is performed on the pre-compression B-mode image $I_1 \in H \times W$ to match the patch size (14×14) of DINOv2. The padded image is fed into the frozen DINOv2 backbone to produce a high-level semantic feature map $F \in \frac{H}{14} \times \frac{W}{14} \times 1024$. To align with the dimension $\frac{H}{8} \times \frac{W}{8} \times C$ of RAFT's context branch output, i.e., $F_{context} \in \frac{H}{8} \times \frac{W}{8} \times C$, bilinear interpolation is implemented on the DINOv2 features. A subsequent 1×1 convolution reduces channel dimensions, yielding $F_{DINO} \in \frac{H}{8} \times \frac{W}{8} \times C$.

### C. MoE-based Feature Fusion Module

Since $F_{context}$ and DINOv2 encode complementary but distinct information, direct concatenation of the two may not be optimal for RAFT-based motion estimation. Specifically, $F_{context}$ captures task-specific contextual information from pre-compression ultrasound frames, whereas DINOv2 provides global semantics. To adaptively fuse these heterogeneous representations, we introduce an MoE-based feature fusion module (Fig. 1).

This module only operates on the context branch. It produces pixel-level routing weights via a softmax activation while employing four independent experts to achieve spatially adaptive feature fusion.

The concatenated joint feature $F_{cat}$ is constructed as $F_{cat} = \text{Concat}(F_{context}, F_{DINO})$. Four experts with identical network structures but independent gating parameters are defined as:

$$E_i(F_{cat}) = F_{context} + \sigma\big(Conv_{1\times1}(F_{cat})\big) \odot F_{DINO}, \quad i = 1,2,3,4 \tag{2}$$

where $Conv_{1\times1}$ is a 1×1 convolution layer followed by a sigmoid activation and $\odot$ represents element-wise multiplication. Low-level contextual features $F_{context}$ and high-level DINOv2 semantics $F_{DINO}$ are combined in a residual manner, with an independent gating convolution followed by a sigmoid activation to control the injection strength of global semantics for regions with distinct deformation patterns.

The routing head applies a convolutional layer over $F_{cat}$ to generate pixel-level routing logits. A temperature factor T scales these logits before Softmax activation, yielding pixel-level gating weights $\{g_1, g_2, g_3, g_4\}$. Then, the fused features are obtained through weighted summation:

$$F_{fusion} = \sum_{i=1}^{N_{experts}} g_i \cdot E_i(F_{cat}) \tag{3}$$

$F_{fusion}$ is then fed into the iterative update module. The pixel-adaptive injection of DINOv2 semantic features provides global contextual representations that complement local matching information for lateral motion estimation.

### D. Unsupervised Loss Function

We adopt a fully unsupervised training framework, where network optimization relies only on ultrasound intensity consistency and tissue-mechanical priors without ground-truth. The total loss is a weighted combination of four terms: similarity loss $\mathcal{L}_D$, strain-smoothness regularization $\mathcal{L}_S$, physical constraint on the EPR $\mathcal{L}_V$ [5], and MoE routing auxiliary loss $\mathcal{L}_{moe}$ [62]:

$$\mathcal{L}_{total} = \mathcal{L}_D + \alpha\mathcal{L}_S + \eta\mathcal{L}_V + \lambda_{moe}\mathcal{L}_{moe} \tag{4}$$

where $\alpha$, $\eta$, and $\lambda_{moe}$ are hyper-parameters (Table I).

The similarity loss is defined by the $L_1$ norm over all pixels of the image grid with size $N \times N$ [63]:

$$\mathcal{L}_D = \left\|\left(I_1 - \hat{I}_2\right)\right\|_{1(N\times N)} \tag{5}$$

where $I_1$ denotes the pre-deformation frame, $I_2$ is the post-deformation frame, $\tilde{\boldsymbol{u}}$ is the predicted displacement field, and $\hat{I}_2$ represents the warped post-deformation image after applying $\tilde{\boldsymbol{u}}$.

The strain-smoothness regularization term $\mathcal{L}_S$ encourages spatially smooth strain estimates, which consists of two components $\mathcal{L}_{s1}$ and $\mathcal{L}_{s2}$ [63]:

$$\mathcal{L}_S = \mathcal{L}_{s1} + \gamma\mathcal{L}_{s2} \tag{6}$$

where $\gamma$ is a weighting coefficient. The term $\mathcal{L}_{s1}$ penalizes the deviation of axial and lateral strain from their global mean values:

$$\mathcal{L}_{s1} = \|\tilde{\varepsilon}_{\text{aa}} - \langle\tilde{\varepsilon}_{\text{aa}}\rangle\|_1 + \|\tilde{\varepsilon}_{ll} - \langle\tilde{\varepsilon}_{\text{ll}}\rangle\|_1 \tag{7}$$

where $\langle\cdot\rangle$ denotes the global pixel-wise mean operator over the whole image. $\mathcal{L}_{s2}$ imposes smoothness on the spatial gradients of strain:

$$\mathcal{L}_{s2} = \left\|\frac{\partial\tilde{\varepsilon}_{\text{aa}}}{\partial a}\right\|_1 + \beta\left\|\frac{\partial\tilde{\varepsilon}_{\text{aa}}}{\partial l}\right\|_1 + 0.5\left\|\frac{\partial\tilde{\varepsilon}_{\text{ll}}}{\partial a}\right\|_1 + 0.5\beta\left\|\frac{\partial\tilde{\varepsilon}_{\text{ll}}}{\partial l}\right\|_1 \tag{8}$$

where $\beta$ weights the relative strength of lateral gradient regularization; $\partial/\partial a$ and $\partial/\partial l$ denote spatial partial derivative along axial and lateral directions.

We follow [5] to incorporate the physical constraint $\mathcal{L}_V$. This prior restricts the EPR within the physiologically reasonable range [0.1, 0.6] and enforces spatial smoothness of the estimated EPR map. The relevant hyperparameters including $V_{emin}$, $V_{emax}$, $\lambda_{vs}$ and $\beta$ are listed in Table I.

TABLE I

HYPERPARAMETERS VALUES AND DESCRIPTIONS. THE NUMBERS INSIDE THE PARENTHESIS INDICATE THE VALUES USED FOR *IN-VIVO* DATA.

| Parameter | Description | Value |
|---|---|---|
| $V_{\text{emin}}$ | minimum range of accepted EPR | 0.1 |
| $V_{\text{emax}}$ | maximum range of accepted EPR | 0.6 |
| $\beta$ | related to ratio of axial to lateral distance | 0.1 |
| $\gamma$ | weight of second-order smoothness | 8 |
| $\alpha$ | weight of smoothness regularizer | 0.1 |
| $\lambda_{vs}$ | smoothness weight of EPR | 0.25 |
| $\eta$ | weight of physic loss | $10^{-3}(10^{-7})$ |
| $w_b$ | weight of router_balance | 5 |
| $w_e$ | weight of router_entropy | 1 |
| $\lambda_{moe}$ | weight of MoE loss | 0.001 |

To mitigate MoE routing collapse, where most pixels persistently activate only one expert branch, we construct a two-component regularization loss $\mathcal{L}_{\text{moe}}$ combining load-balancing $\mathcal{L}_{balance}$ and routing-entropy-maximization terms $\mathcal{L}_{entropy}$ [62]:

$$\mathcal{L}_{moe} = w_b \cdot \mathcal{L}_{balance} + w_e \cdot \mathcal{L}_{entropy} \tag{9}$$

$$usage_i = \mathbb{E}_{N,H,W}[g_i] \tag{10}$$

where $g_i$ is the routing probability of pixel assigned to expert i; $\mathbb{E}_{N,H,W}[\cdot]$ denotes expectation over spatial dimensions. $\mathcal{L}_{balance}$ forces all experts to be activated with similar probability:

$$\mathcal{L}_{balance} = \sum_{i=1}^{N_{experts}} \eta_i \cdot \left(usage_i - \frac{1}{N_{experts}}\right)^2 \tag{11}$$

where $N_{experts}$ is total expert number, and $\eta_i$ is weight coefficient for expert i. The entropy of routing distribution $H(g)$ is defined as:

$$H(g) = -\sum_{i=1}^{N_{experts}} g_i \cdot \log(g_i + \epsilon), \quad \epsilon = 10^{-8} \tag{12}$$

The entropy maximization loss $\mathcal{L}_{entropy}$ encourages the routing distribution close to maximum entropy:

$$\mathcal{L}_{entropy} = \max H - \mathbb{E}_{N,H,W}[H(g)] \tag{13}$$

where $\max H$ is the theoretical maximum entropy for uniform routing probability.

## IV. EXPERIMENTS

### A. Datasets

Three datasets are used in this study: a simulated dataset and two *in-vivo* datasets.

The simulation dataset [32] was downloaded from *data.sonography.ai*. The Field-II simulator generated pre- and post-deformation RF signals for 24 phantoms, each with 10 different scatterer distributions (240 configurations total). Each phantom contained one or two circular lesions with Young's modulus randomly sampled from 40-60 kPa, which were embedded in homogeneous background tissue (18-23 kPa). Applied strain varied from 0% to 4.5% in steps of 0.5%, and ground-truth displacement fields were solved by the finite-element method (FEM).

Image pairs are constructed using frames before and after deformation under different applied strains, producing 90 pairs per phantom. The image dimensions are 2048 × 256. The axial sample spacing is 0.015 mm; lateral line spacing is 0.13 mm. The dataset is split into 2,574 training pairs and 630 validation samples with an approximate ratio of 4:1.

The first *in-vivo* dataset was acquired from patients with breast tumors under the approval of the Institutional Review Board (IRB) of Beijing Tsinghua Changgung Hospital, and written informed consent was obtained from all enrolled patients. Data collection was performed using a Vantage 256 ultrasound system (Verasonics Inc., Kirkland, WA, USA) equipped with an L10-5 linear-array probe. Manual tissue compression was applied to produce tissue strain. The RF data were acquired with a 6.25 MHz center frequency and a 25 MHz sampling frequency. The axial sample spacing and lateral line spacing are 0.03 mm and 0.3 mm, respectively. This dataset is used as the training and validation dataset.

The second *in-vivo* dataset was collected with a Siemens Elegra ultrasound scanner equipped with a 7.5 MHz high-frequency linear-array probe [64, 65] from patients with pathologically confirmed breast lesions. Manual compression generated tissue strain of 1%-1.5%. The RF sampling frequency was 36 MHz, with axial and lateral sample spacings of 0.021 mm and 0.19 mm. Data acquisition was approved by the IRB of the University of Kansas Medical Center, and informed patient consent was obtained. Each image pair is formed by randomly picking two frames from continuous recordings, yielding 14 *in-vivo* image pairs for evaluation. This dataset is used as the external test dataset.

### B. Training Details

For data preprocessing, we adopt a multi-modal input strategy [32]. RF signals, RF envelopes and B-mode images are concatenated into a three-channel input tensor. All modalities are normalized to [0,1]. Zero-mean Gaussian noise with a standard deviation of 0.05 is injected as data augmentation.

Following standard transfer-learning practice for vision foundation models, the DINOv2 backbone is frozen throughout training.

The AdamW optimizer is used to mitigate overfitting via weight decay. The base learning rate is set to $1\times10^{-4}$ with a weight-decay coefficient of $5\times10^{-5}$. The OneCycleLR scheduler runs for 150,000 total steps: the learning rate linearly ramps up for the initial 5% of iterations and then linearly anneals for the remaining steps to balance rapid initial fitting and fine late-stage convergence. FP16 mixed precision with gradient scaling reduces GPU memory consumption. The batch size is fixed to 1 for high-resolution 2048 × 256 ultrasound inputs. DataParallel is utilized for multi-GPU training.

### C. Quantitative Metrics

The output of DinoMUSE is the axial and lateral displacements, while strains are of interest in USE. Therefore, Savitzky–Golay digital differentiator is used to calculate the axial, lateral, axial-shear and lateral-shear strains from displacements [66].

For quantitative evaluation, five metrics are adopted: normalized root-mean-square error (NRMSE) of displacement or strain estimation [32], mean-squared error (MSE) of displacement or strain estimation, mean absolute error (MAE) of displacement or strain estimation [5], SNR of strain image, and CNR of strain image [29]:

$$NRMSE = \sqrt{\mathbb{E}\left[\left(\frac{\phi_{\text{pred}} - \phi_{\text{gt}}}{\phi_{\text{gt}}}\right)^2\right]}, MSE = \mathbb{E}\left[\left(\phi_{\text{pred}} - \phi_{\text{gt}}\right)^2\right] \quad (14)$$

$$MAE = \mathbb{E}\left[\left|\phi_{\text{pred}} - \phi_{\text{gt}}\right|\right] \quad (15)$$

$$\text{SNR} = \frac{|\mu_T|}{\sigma_T}, \text{CNR} = \sqrt{\frac{2\cdot(\mu_B - \mu_T)^2}{\sigma_B^2 + \sigma_T^2}} \quad (16)$$

where $\phi_{\text{pred}}$ and $\phi_{\text{gt}}$ are the predicted and ground-truth fields (either displacement or strain or EPR value), $\mathbb{E}[\cdot]$ represents the expectation operator calculated over all pixels inside the region of interest (ROI). $\mu_T$ and $\sigma_T$ are the mean value and standard deviation of strain values within the lesion ROI, $\mu_B$ and $\sigma_B$ are the mean value and standard deviation of strain values within the background tissue ROI.

We compared the proposed method, DinoMUSE with NCC [22], GLUE [29], RAFT [49] and sPICTURE [5] on the same datasets. NCC and GLUE operate on RF data, whereas RAFT, sPICTURE, and DinoMUSE employ three-channel inputs. Hyperparameters for each competing algorithm are tuned to yield optimal performance, ensuring a fair comparison. For NCC, we set axial and lateral window sizes to 27 axial samples × 7 lateral lines (0.52 mm × 0.875 mm) for simulation data and 41 axial samples × 11 lateral lines (0.79 mm × 2.20 mm) for *in-vivo* data. For GLUE, the regularization coefficients were set following the recommended values in [29], more specifically, $\alpha_1 = 5$, $\alpha_2 = 1$, $\beta_1 = 5$, $\beta_2 = 1$, and $\alpha_{DP} = 0.2$. For *in-vivo* data, these parameters are adjusted as $\alpha_1 = 0.05$, $\alpha_2 = 0.01$, $\beta_1 = 0.05$, $\beta_2 = 0.01$, and $\alpha_{DP} = 0.2$. For a fair comparison

TABLE II

QUANTITATIVE RESULTS OF ESTIMATED AXIAL AND LATERAL DISPLACEMENTS AND EPR FOR THE SIMULATION TEST DATASET (MEAN ± SD, 630 IMAGE PAIRS)

| Method | Axial displacement | | | Lateral displacement | | | EPR | | | Runtime |
|---|---|---|---|---|---|---|---|---|---|---|
| | NRMSE ↓ | MSE ↓ | MAE ↓ | NRMSE ↓ | MSE ↓ | MAE ↓ | NRMSE ↓ | MSE ↓ | MAE ↓ | (sec) ↓ |
| NCC | 0.173±0.065 | 2.919±3.519 | 1.148±0.859 | 997.34±1233.38 | 0.08±0.07 | 0.186±0.098 | $2.33\times10^3$±$2.58\times10^3$ | $2.64\times10^6$±$5.90\times10^6$ | 40.657±46.160 | 32.2584 |
| GLUE | 0.023±0.008 | 0.060±0.064 | 0.183±0.111 | 2109.51±2375.63 | 0.12±0.11 | 0.190±0.099 | 24.77±260.37 | $3.87\times10^3$±$8.24\times10^3$ | 0.191±0.158 | 13.1832 |
| RAFT | 0.019±0.005 | 0.048±0.048 | 0.178±0.102 | 742.98±1243.00 | 0.04±0.03 | 0.131±0.065 | 0.37±0.05 | 0.24±0.55 | 0.101±0.008 | **0.1790** |
| sPICTURE | 0.019±0.005 | 0.047±0.049 | 0.174±0.104 | 615.90±785.47 | 0.05±0.05 | 0.146±0.076 | 0.46±0.07 | 0.26±0.54 | 0.132±0.017 | 0.2204 |
| DinoMUSE | **0.017±0.004** | **0.046±0.046** | **0.172±0.100** | **471.41±754.89** | **0.02±0.02** | **0.101±0.045** | **0.28±0.05** | **0.23±0.55** | **0.079±0.030** | 0.1956 |

TABLE III

QUANTITATIVE RESULTS OF ESTIMATED AXIAL AND LATERAL STRAINS FOR THE SIMULATION TEST DATASET (MEAN ± SD, 630 IMAGE PAIRS). CNR AND SNR WERE EVALUATED ON A SUBSET OF 90 IMAGE PAIRS FROM ONE PHANTOM.

| Method | Axial Strain | | | | | Lateral Strain | | | | |
|---|---|---|---|---|---|---|---|---|---|---|
| | NRMSE↓ | MSE↓ | MAE↓ | SNR↑ | CNR↑ | NRMSE↓ | MSE↓ | MAE↓ | SNR↑ | CNR↑ |
| NCC | 0.54±0.53 | 14.06±39.43 | 0.535±0.314 | 4.61±1.85 | 2.05±0.63 | 1.35±0.41 | 30.43±48.55 | 0.93±0.56 | 0.99±0.37 | 0.69±0.33 |
| GLUE | 0.44±0.73 | 0.60±0.50 | 0.145±0.059 | 7.36±3.54 | 4.05±1.79 | 0.58±0.13 | 2.39±2.09 | 0.35±0.17 | 4.04±2.13 | 2.94±1.33 |
| RAFT | 0.42±0.80 | 0.20±0.19 | 0.062±0.026 | 35.87±16.29 | 20.94±6.92 | 0.36±0.03 | 1.36±1.23 | 0.24±0.12 | 16.99±6.65 | 4.22±2.27 |
| sPICTURE | **0.40±0.76** | 0.20±0.18 | 0.073±0.024 | 27.93±11.42 | 14.75±5.40 | 0.44±0.07 | 2.85±2.84 | 0.33±0.19 | 9.87±2.56 | 10.59±4.29 |
| DinoMUSE | 0.42±0.83 | **0.18±0.18** | **0.055±0.023** | **50.76±25.91** | **28.08±9.36** | **0.27±0.04** | **0.47±0.36** | **0.16±0.06** | **46.62±17.93** | **17.81±5.92** |
| Method | Axial Shear Strain | | | | | Lateral Shear Strain | | | | |
| | NRMSE↓ | MSE↓ | MAE↓ | SNR↑ | CNR↑ | NRMSE↓ | MSE↓ | MAE↓ | SNR↑ | CNR↑ |
| NCC | 6781.2±11988.9 | 356.87±2695.46 | 1.31±1.00 | 0.09±0.06 | 1.12±0.48 | 789.07±621.18 | 0.43±0.55 | 0.110±0.058 | 0.16±0.11 | 0.264±0.198 |
| GLUE | 2208.4±1730.2 | 13.74±11.32 | 0.57±0.27 | 0.12±0.10 | 2.16±0.76 | 865.86±563.36 | 1.13±0.94 | 0.164±0.078 | 0.15±0.10 | 0.317±0.229 |
| RAFT | 2250.2±2073.2 | 5.58±4.51 | 0.33±0.14 | 0.55±0.53 | 5.02±1.97 | 434.16±277.57 | 0.11±0.10 | 0.073±0.037 | 0.31±0.22 | 0.451±0.305 |
| sPICTURE | **1771.4±1469.6** | **4.54±3.62** | 0.31±0.13 | 0.25±0.23 | 4.79±1.39 | 295.30±183.36 | 0.06±0.05 | 0.053±0.026 | 0.29±0.24 | 0.517±0.381 |
| DinoMUSE | 2098.3±1822.1 | 4.77±3.94 | **0.30±0.13** | **0.62±0.61** | **6.09±1.96** | **129.95±89.25** | **0.02±0.02** | **0.033±0.015** | **0.80±0.54** | **0.967±0.644** |

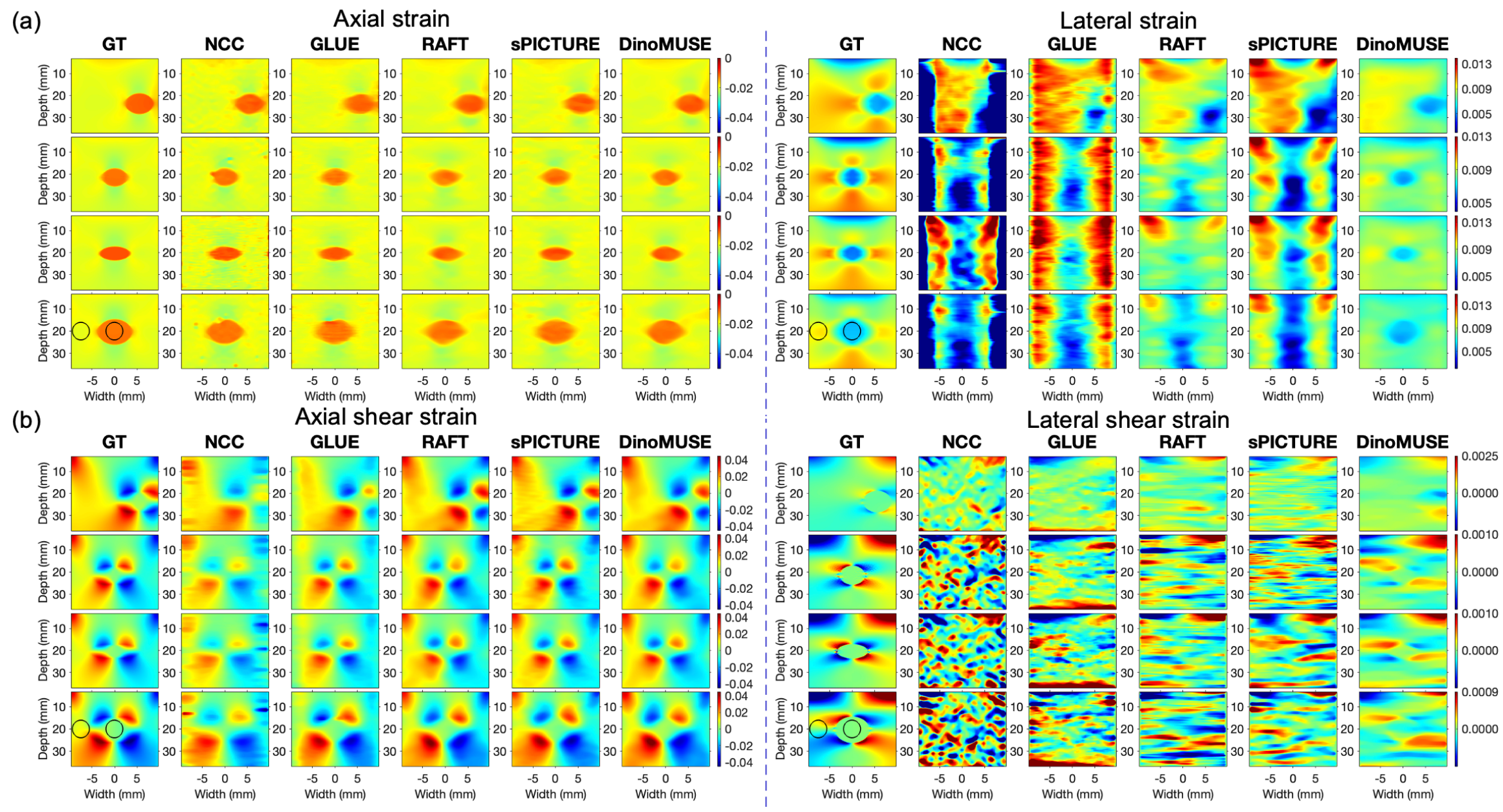

Fig. 2. Estimated axial and lateral strain images of four simulated phantoms. (a) Axial (left) and lateral (right) strains. (b) Axial-shear (left) and lateral-shear (right) strains. Black circles denote the ROIs in the target region and the background region selected for CNR and SNR calculation.

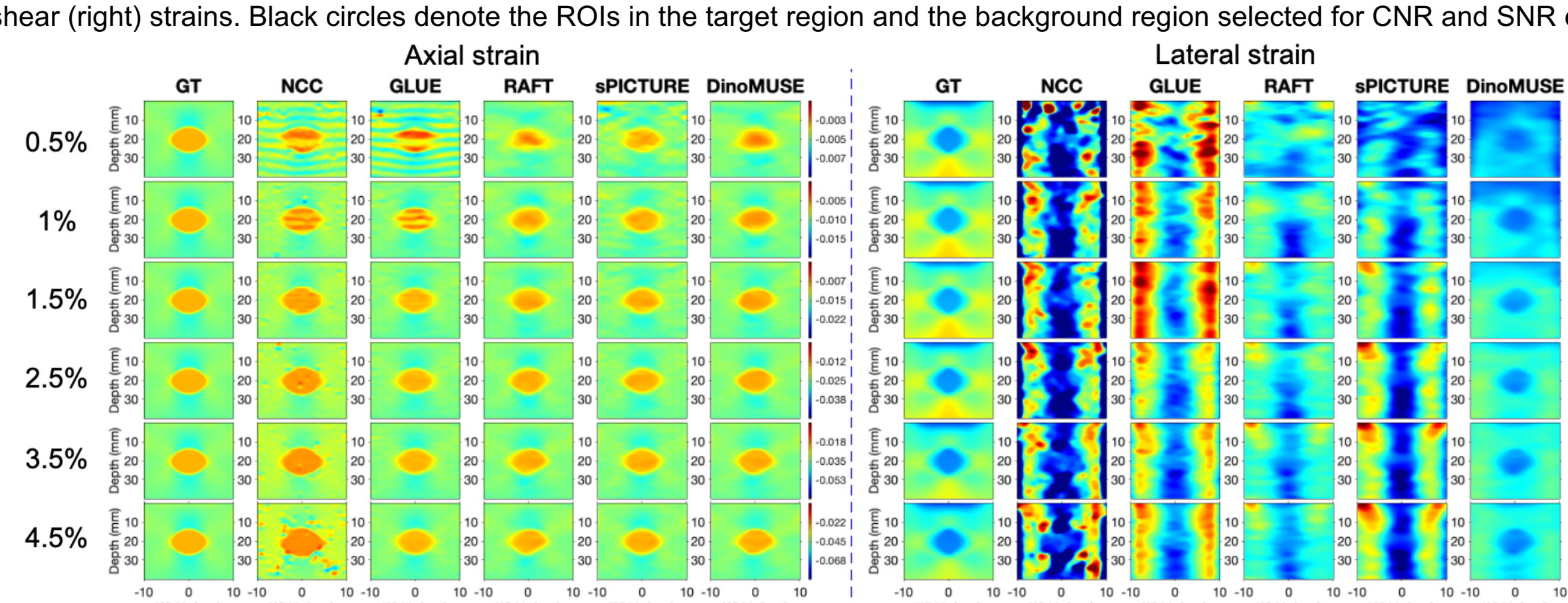

Fig. 3. Estimated axial (left) and lateral (right) strain images of phantom 4 at different applied strain levels

across all competing approaches, displacements from all algorithms were post-processed using the same linear-fit Savitzky-Golay digital differentiator [66] for strain computation, with filter length set to 1 mm for axial differentiation and 2 mm for lateral differentiation.

## D. Simulation Results

Table II summarizes results of axial and lateral displacement fields for all five algorithms. All methods achieve low axial displacement errors, especially deep-learning-based ones. Our method obtains the best performance with an axial NRMSE of 0.0167. For the lateral direction, our DinoMUSE achieves a lateral MSE of only 0.02, corresponding to a 60 % reduction compared with sPICTURE.

Table III reports quantitative metrics for axial, lateral, axial-shear and lateral-shear strains, and Fig. 2 visualizes the corresponding strain images. Model differences are minor for axial strain, with our method achieving the best MSE, MAE, CNR and SNR. In contrast, larger performance gaps are observed for lateral strain. NCC and GLUE generate heavily distorted strain images; RAFT can roughly distinguish lesions; sPICTURE suppresses part of the estimation errors via EPR constraints yet suffers from insufficient lesion-to-background contrast. Our method yields a low lateral strain MSE of 0.47 with markedly higher CNR and SNR.

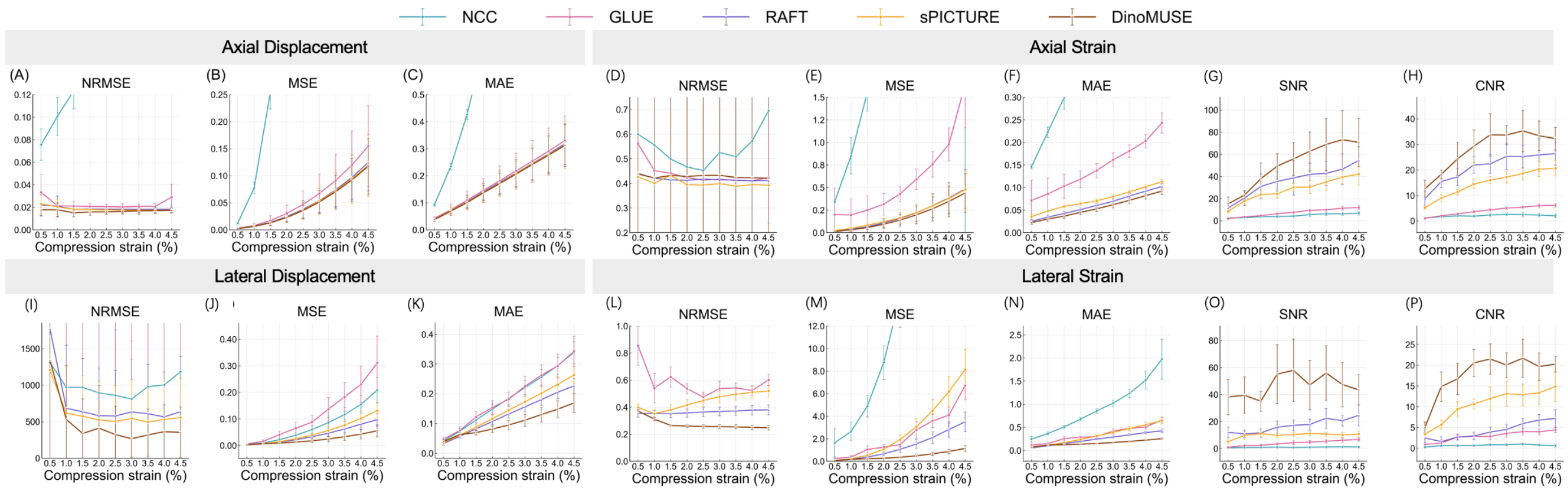

Fig. 4. Performance analysis of all algorithms at different applied strain levels (MEAN ± SD, 630 image pairs)

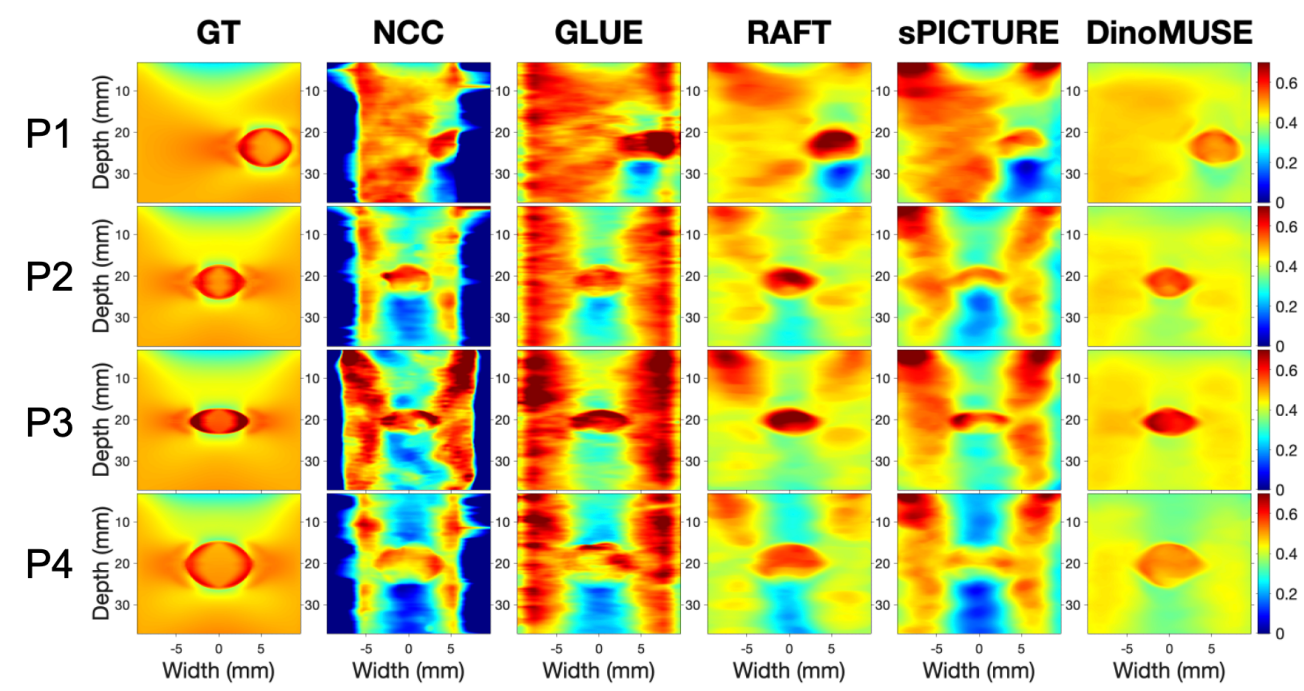

Fig. 5. EPR of four phantoms obtained by different algorithms

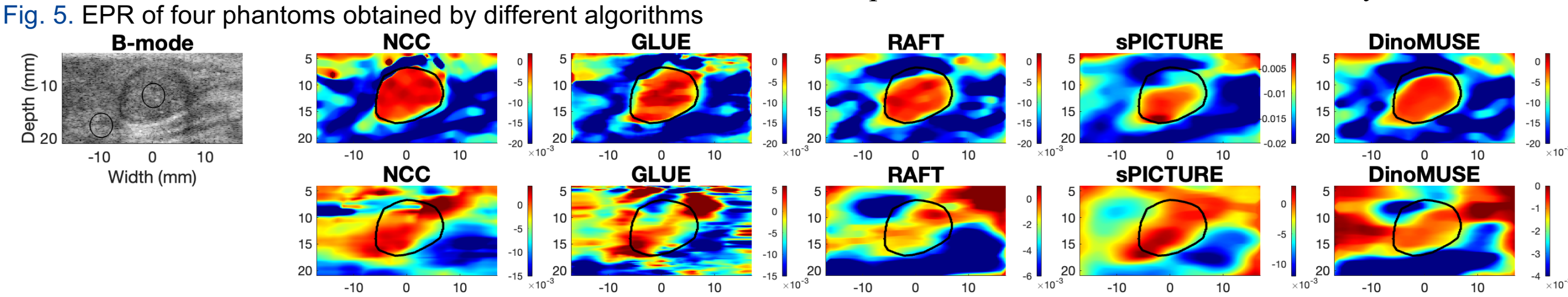

Fig. 6. B-mode image of the lesion and the corresponding axial (up) and lateral (down) strain images estimated by all algorithms. Black solid circles in the B-mode image denote the ROIs in the target region and the background region selected for CNR and SNR calculation. Black contours trace lesion margins on each strain map.

Figure 3 displays axial and lateral strain images under different strain ratios. Figure 4 plots curves of axial/lateral displacement NRMSE as well as strain MSE, CNR and SNR under compression strains ranging from 0.5% to 4.5%. Our method maintains the lowest lateral-strain error throughout the strain range. Particularly under high-strain conditions, competing methods exhibit increased lateral-strain MSE, while DinoMUSE achieves superior robustness.

Figure 5 presents visual comparisons of EPR across four phantoms for different algorithms. The ground-truth EPR maps show a smooth tissue distribution and clear lesion boundaries. NCC and GLUE produce heavily distorted EPR maps with severe spurious fluctuations. RAFT and sPICTURE reduce errors yet still introduce obvious false variations in homogeneous backgrounds. Our DinoMUSE yields EPR maps with minimal spurious fluctuations, well-preserved lesion boundaries, and the best visual consistency with the ground truth. Consistent with these visual observations, Table II reports the EPR metrics, where DinoMUSE obtains the lowest error against ground-truth EPR.

These results demonstrate that the proposed method effectively improves the estimation accuracy of lateral strain, while maintaining accurate axial strain estimation.

## *E. In-vivo Results*

For the *in-vivo* external test dataset, since ground-truth displacement fields are unavailable, only SNR and CNR are adopted as quantitative metrics for strain imaging. Quantitative results are summarized in Table IV, and the corresponding axial and lateral strain images are presented in Fig. 6. Our DinoMUSE yields the clearest lesion appearance and achieves the optimal SNR and CNR in axial and lateral directions.

TABLE IV

QUANTITATIVE RESULTS OF ESTIMATED AXIAL AND LATERAL STRAINS FOR THE *IN-VIVO* DATASET (MEAN ± SD, 14 IMAGE PAIRS)

| Method | Axial strain | | Lateral strain | |
|---|---|---|---|---|
| | SNR↑ | CNR↑ | SNR↑ | CNR↑ |
| NCC | 0.62±0.36 | 1.29±0.78 | 0.26±0.14 | 0.47±0.26 |
| GLUE | 0.16±0.15 | 0.88±0.70 | 0.35±0.29 | 0.39±0.30 |
| RAFT | 1.42±0.54 | 2.49±1.16 | 2.38±1.23 | 1.78±0.93 |
| sPICTURE | 4.11±2.85 | 3.32±2.47 | 4.57±3.30 | 2.13±2.15 |
| DinoMUSE | **5.70±3.70** | **4.82±3.01** | **4.61±2.38** | **2.23±1.99** |

## *F. Ablation Studies*

Since DinoMUSE incorporates the EPR physical constraint, ablation studies are conducted on the simulation dataset to dissect the contribution of the EPR prior, DINOv2-guided feature injection, and the MoE-based fusion module.

Results are shown in Table V and VI. Adding only the EPR constraint slightly degrades axial displacement accuracy. Nevertheless, the lateral strain MSE is substantially reduced,

TABLE V

QUANTITATIVE RESULTS OF ESTIMATED AXIAL AND LATERAL DISPLACEMENTS FOR ABLATION STUDY I ON THE SIMULATION DATASET (MEAN ± SD, 630 IMAGE PAIRS)

| Method | EPR | DINOv2 | MoE | Axial displacement | | | Lateral displacement | | | Runtime (sec) ↓ |
|---|---|---|---|---|---|---|---|---|---|---|
| | | | | NRMSE↓ | MSE↓ | MAE↓ | NRMSE↓ | MSE↓ | MAE↓ | |
| RAFT | - | - | - | 0.0190±0.0051 | 0.0483±0.0479 | 0.176±0.102 | 742.98±1243.00 | 0.0399±0.0341 | 0.131±0.065 | 0.1790 |
| RAFT+EPR | √ | - | - | 0.0196±0.0038 | 0.0517±0.0568 | 0.179±0.113 | 733.10±966.23 | 0.0232±0.0239 | 0.110±0.052 | **0.1779** |
| DinoGATE | √ | √ | - | 0.0170±0.0045 | **0.0455±0.0444** | 0.173±0.098 | **444.59±541.52** | 0.0232±0.0208 | 0.102±0.045 | 0.1870 |
| DinoMUSE | √ | √ | √ | **0.0167±0.0037** | 0.0457±0.0460 | **0.172±0.100** | 471.41±754.89 | **0.0226±0.0202** | **0.101±0.045** | 0.1956 |

TABLE VI

QUANTITATIVE RESULTS OF ESTIMATED AXIAL AND LATERAL STRAINS FOR ABLATION STUDY I ON THE SIMULATION DATASET (MEAN ± SD, 630 IMAGE PAIRS). CNR AND SNR WERE EVALUATED ON A SUBSET OF 90 IMAGE PAIRS FROM ONE PHANTOM.

| Method | Axial strain | | | | | Lateral strain | | | | |
|---|---|---|---|---|---|---|---|---|---|---|
| | NRMSE↓ | MSE↓ | MAE↓ | CNR↑ | SNR↑ | NRMSE↓ | MSE↓ | MAE↓ | CNR↑ | SNR↑ |
| RAFT | 0.42±0.80 | 0.20±0.19 | 0.062±0.026 | 20.94±6.92 | 35.87±16.29 | 0.36±0.03 | 1.36±1.23 | 0.24±0.12 | 4.22±2.27 | 16.99±6.65 |
| RAFT+EPR | **0.41±0.79** | 0.20±0.19 | 0.061±0.025 | 21.54±6.73 | 36.79±18.24 | 0.28±0.03 | 0.51±0.52 | **0.15±0.08** | 13.51±3.55 | 38.16±17.41 |
| DinoGATE | 0.42±0.83 | **0.17±0.18** | **0.054±0.023** | 26.76±8.95 | 45.05±17.36 | 0.30±0.05 | 0.48±0.37 | 0.16±0.06 | 15.72±5.82 | 45.54±20.85 |
| DinoMUSE | 0.43±0.83 | 0.18±0.18 | 0.055±0.023 | **28.08±9.36** | **50.76±25.91** | **0.27±0.04** | **0.47±0.36** | 0.16±0.06 | **17.81±5.92** | **46.62±17.93** |

accompanied by a improved visual quality and sharper lesion boundaries.

Building on EPR regularization, the introduction of DINOv2 features yields additional performance gains. The proposed DinoMUSE, equipped with the MoE module, further boosts axial accuracy and achieves a reduction of nearly 36% in lateral displacement NRMSE compared with that of RAFT+EPR. Although DinoMUSE exhibits slight degradation in several individual metrics relative to DinoGATE, it attains the best overall quantitative performance.

## G. DINOv2 Feature Maps After MoE fusion

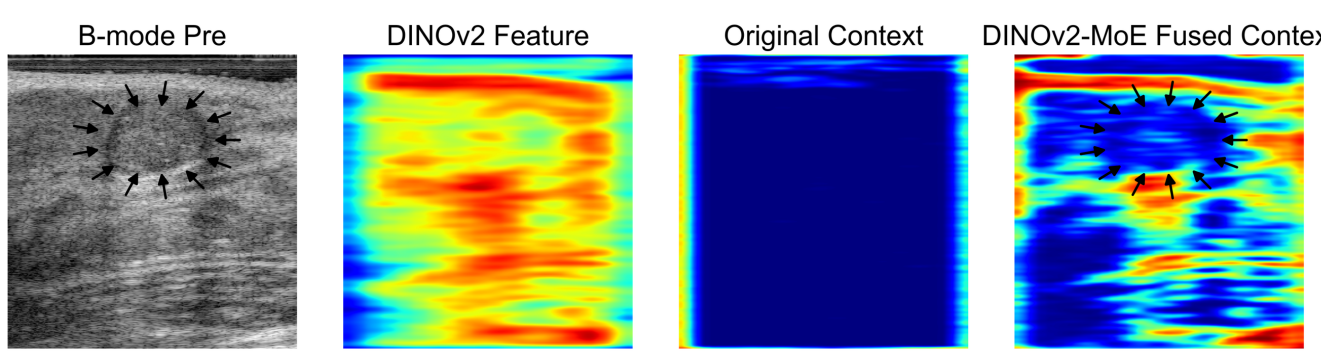

Fig. 7. Qualitative comparison of intermediate context representations with and without DINOv2-MoE fusion. Black arrows outline the inclusion. From left to right: raw B-mode image, DINOv2-feature map, original RAFT context feature map without DINO guidance, and MoE-fused context-feature feature map.

Furthermore, Fig. 7 visualizes intermediate context features to reveal the effect of the MoE context fusion module. The RAFT original context feature map exhibits relatively homogeneous activation and lacks spatial modulation aligned with anatomical structures. After MoE fusion, context features incorporate DINO structural priors, and the response patterns closely match ultrasound anatomy.

Compared with standalone DINO feature maps, MoE-modulated representations show better alignment with B-mode anatomical layouts. This demonstrates that the MoE module adaptively injects DINO semantic priors into the RAFT context branch, amplifying feature responses in regions of interest and supplying more discriminative cues for iterative solving of the motion field, thereby boosting motion estimation performance in USE.

## H. Analysis of MoE-based Feature Fusion Module

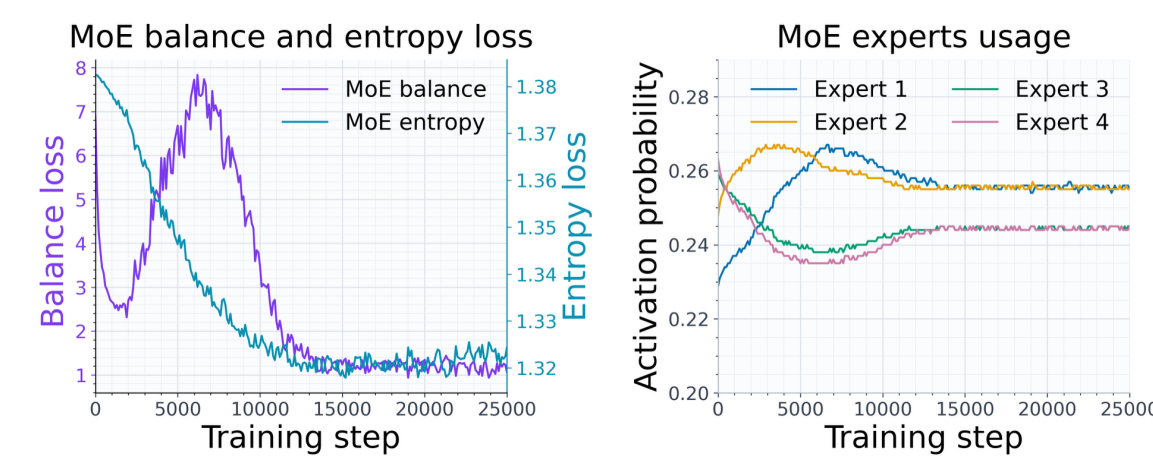

Fig. 8. MoE Loss curves during unsupervised learning with simulation data and MoE experts usage

The MoE training-loss curves, entropy variations and expert-usage visualizations are shown in Fig. 8. At the early training stage, average activation probabilities of the four experts diverge greatly, indicating unstable routing allocation. As training proceeds, constrained by the balance loss, expert-usage ratios gradually converge, effectively avoiding routing collapse. Nevertheless, fully equal activation ratios are not attained upon convergence. Expert 1 and Expert 2 maintain relatively higher global usage while Expert 3 and Expert 4 are activated less frequently, suggesting a spontaneous and moderate expert preference driven by characteristics of the tissue motion estimation task.

Besides, we analyzed the influence of the number of experts.

TABLE VII

QUANTITATIVE RESULTS OF ESTIMATED AXIAL AND LATERAL DISPLACEMENTS FOR ABLATION STUDY II ON THE SIMULATION DATASET (MEAN ± SD, 630 IMAGE PAIRS)

| Method | Number of experts | Axial displacement | | | Lateral displacement | | | Runtime (sec) ↓ |
|---|---|---|---|---|---|---|---|---|
| | | NRMSE↓ | MSE↓ | MAE↓ | NRMSE↓ | MSE↓ | MAE↓ | |
| DinoGATE | 1 | 0.0170±0.0045 | **0.0455±0.0444** | 0.173±0.098 | **444.59±541.52** | 0.0232±0.0208 | 0.102±0.045 | **0.1870** |
| DinoMoE2 | 2 | 0.0172±0.0053 | 0.0467±0.0455 | 0.174±0.100 | 590.85±818.98 | 0.0252±0.0230 | 0.106±0.050 | 0.1875 |
| DinoMoE3 | 3 | 0.0172±0.0038 | 0.0462±0.0460 | 0.173±0.100 | 510.68±649.83 | 0.0243±0.0236 | 0.102±0.053 | 0.1910 |
| DinoMUSE | 4 | **0.0167±0.0037** | 0.0457±0.0460 | **0.172±0.100** | 471.41±754.89 | **0.0226±0.0202** | **0.101±0.045** | 0.1956 |

TABLE VIII

QUANTITATIVE RESULTS OF ESTIMATED AXIAL AND LATERAL STRAINS FOR ABLATION STUDY II ON THE SIMULATION DATASET (MEAN ± SD, 630 IMAGE PAIRS). CNR AND SNR WERE EVALUATED ON A SUBSET OF 90 IMAGE PAIRS FROM ONE PHANTOM.

| Method | Axial strain | | | | | Lateral strain | | | | |
|---|---|---|---|---|---|---|---|---|---|---|
| | NRMSE↓ | MSE↓ | MAE↓ | CNR↑ | SNR↑ | NRMSE↓ | MSE↓ | MAE↓ | CNR↑ | SNR↑ |
| DinoGATE | **0.413±0.831** | 0.175±0.177 | **0.0544±0.0225** | 26.76±8.95 | 45.05±17.36 | 0.30±0.05 | 0.48±0.37 | 0.1642±0.0550 | 15.72±5.82 | 45.54±20.85 |
| DinoMoE2 | 0.419±0.813 | **0.173±0.175** | 0.0547±0.0226 | 26.15±8.27 | 41.42±17.88 | 0.28±0.03 | 0.52±0.45 | 0.1562±0.0621 | 15.70±4.79 | 37.68±13.44 |
| DinoMoE3 | 0.428±0.832 | 0.175±0.176 | 0.0548±0.0221 | 26.17±8.74 | 41.39±18.30 | 0.29±0.03 | 0.49±0.43 | **0.1559±0.0652** | 14.56±5.78 | 40.70±16.46 |
| DinoMUSE | 0.428±0.833 | 0.175±0.178 | 0.0549±0.0233 | **28.08±9.36** | **50.76±25.91** | **0.27±0.04** | **0.47±0.36** | 0.1589±0.0550 | **17.81±5.92** | **46.62±17.93** |

The results in Table VII and VIII show that estimation performance improves as the number of experts increases. This observation is consistent with the capacity-bottleneck phenomenon reported in GShard [67]. When the number of experts is insufficient, the model capacity cannot adequately accommodate diverse deformation modes and task interference. Once a critical threshold is crossed, the bottleneck is alleviated and substantial performance gains can be obtained. Nevertheless, continuously increasing the number of experts may result in a larger model capacity, increased training complexity, and higher inference overhead. Balancing estimation accuracy and computational cost, we adopt the four-expert DinoMUSE as our final configuration.

### *I. Limitations and Future Work*

This study has several limitations. First, our *in-vivo* data only cover human breast lesions; multi-organ and multi-site datasets for liver, thyroid and other tissues are absent. The cross-organ and cross-scanner generalization capacity requires further validation, though simulation-based experiments have already validated the underlying motion-estimation capability. Second, leveraging DINOv2 features introduces additional computational cost and increases overall inference latency. Future work will explore strategies to mitigate this computational burden. Further research is also needed to investigate the clinical utility of the proposed method.

## V. Conclusion

Targeting the well-known bottleneck of poor lateral motion estimation caused by sparse lateral sampling in ultrasound elastography, this work presents a DINO-guided mixture-of-experts framework for unsupervised motion estimation in USE. Experiments on both simulated and *in-vivo* datasets demonstrate that our model achieves substantial improvements in lateral displacement and strain estimation compared with traditional algorithms (NCC and GLUE) and deep-learning-based methods (RAFT and sPICTURE), while maintaining competitive axial displacement and strain estimation precision.

## Acknowledgments

The authors would like to thank Prof. Timothy Hall from the University of Wisconsin–Madison, Madison, WI, USA, and Prof. Jingfeng Jiang from Michigan Technological University, Houghton, MI, USA, for providing the *in-vivo* data.